# Angular momenta of even-even fragments in the neutronless fission of $^{252}$Cf

Ş. Mişicu,[1] A. Săndulescu,[1,2] G. M. Ter-Akopian,[3] and W. Greiner[2]
[1]*National Institute for Nuclear Physics, Bucharest, P.O. Box MG6, Romania*
[2]*Institut für Theoretische Physik der J.W. Goethe Universität, Frankfurt am Main, Germany*
[3]*Flerov Laboratory for Nuclear Reactions, Joint Institute for Nuclear Research, Dubna, RU-141980, Russia*

The recent advent of experimental techniques in which the dynamical characteristics of fission fragments are determined more accurately prompted us to investigate the angular momentum acquired by fragments in a model which describes the cold (neutronless) fission of $^{252}$Cf as the decay of a giant nuclear molecule. The molecular configuration is a consequence of the interplay between the attractive nuclear part and the repulsive Coulomb+nuclear forces. The basic idea of the present approach is to separate the radial (fission) modes describing the decay of the molecule from the modes associated to transversal vibrations (bending) of the fragments. The distance between the centers of the two fragments is fixed by the requirement that the energy released in the fission reaction $Q$ equals the sum of quantum zero energies of radial and transversal modes and the total excitation energy $E^*$. Using a semiclassical coupled channel formalism we computed the additional angular momenta acquired by the fragments during their postscission motion, and found that the Coulomb excitation accounts for less than 10% of the final spins. [S0556-2813(99)02609-6]

## I. INTRODUCTION

The knowledge of the angular momentum distribution of primary fission fragments is important because it provides information on the fissioning system at the scission point. The experimental and theoretical investigations carried out in the past on the $\gamma$ rays assumed that the mechanism of angular momentum formation may be divided in two stages [1–4]. In the first stage, which takes place at the scission configuration, the fragments acquire angular momentum due to the excitation of collective transversal degrees of freedom induced by the interplay between the nuclear and Coulomb forces. In the second stage an additional angular momentum will come up from the mutual Coulomb excitation. In this paper we deal with both stages, for the limiting case of cold fission, i.e., when the total excitation energy does not exceed the neutron emission threshold.

Recently it has been advocated by us [5], based on the concept of *nuclear molecule* [6], that for fragments emitted with almost no excitation energy, a molecular vibrational spectrum will show up as a consequence of small nonaxial fluctuations at scission. An equilibrium position in the fragment-fragment distance coordinate can be achieved if the interplay between the Coulomb and the repulsive nuclear core on one hand and the attractive nuclear part on the other hand will produce a potential bag. If such a pocket is not too shallow, dipole oscillations of the relative coordinate and rotational vibrations like bending and wriggling can occur [6]. These last two excitations take place perpendicularly to the fission axis and were predicted long time ago by Nix and Swiatecki [7]. There were also predicted other rotational modes like twisting and tilting [8] but we shall not treat them in the present work.

In last time experimental data on the spin distribution of fission fragments were made available, especially through the use of large arrays of high-resolution gamma detectors. The first data on rotational states population in cold fission of $^{252}$Cf were presented in Ref. [9]. Recent measurements of the multiple $\gamma$ rays emitted by different pairs of fragments formed in the spontaneous fission of $^{252}$Cf were analyzed using a two-dimensional matrix of $\gamma-\gamma$ coincidences [10]. From here it was possible to extract the average angular momentum for the primary fission fragments as a function of the neutron multiplicity for the Mo-Ba and Zr-Ce charge splits of $^{252}$Cf. The fission mode corresponding to zero neutrons emitted was supposed to be of adiabatic nature in view of the smaller excitation energy involved in it [11].

In this paper we foccus on the study of the neutronless case employing a model which describes the fissioning system at the scission point as two coupled one-dimensional oscillators performing only bending and wriggling vibrations around the equilibrium position which is chosen to correspond to both fragments having their symmetry axes aligned. The stiffness constant is derived from the expansion of the heavy-ion potential up to second order in the angular deviation. For a fixed amount of total excitation energy one employs a different set of deformation parameters, provided the whole excitation energy is stored in deformation. In this way we were able to compute the average angular momentum of the fragments at scission for zero or very small excitation energies.

Before discussing the obtained results in the light of the most recent reported experimental data we compute the change in angular momentum due to the Coulomb excitation (coulex) using a semiclassical model which includes also the shape dynamics.

## II. GIANT MOLECULE SCENARIO

In order to describe the molecular state formed at scission we adopt an improved version of the receipt proposed by Rasmussen *et al.* [2]. The present formalism is extended to the case of two deformed nuclei, with finite-size effects and diffusivity taken into account. Also care has been taken for the conservation of total angular momentum.

The interaction between the two deformed nuclei $V(\mathbf{R})$ is the sum of a short-range nuclear interaction $V_N(\mathbf{R})$ and the long-range Coulomb $V_C(\mathbf{R})$ parts. It can be calculated as the double folding integral of ground state one-body densities $\rho_{1(2)}(\mathbf{r})$ of heavy ions as follows:

$$V(\mathbf{R}) = \int d\mathbf{r}_1 \int d\mathbf{r}_2 \, \rho_1(\mathbf{r}_1)\rho_2(\mathbf{r}_2) v(s). \quad (1)$$

In previous papers we employed the M3Y *NN* effective interaction for the nuclear part of $v$ [12–14] and we computed the WKB penetrabilities for the binary and ternary cold fission of $^{252}$Cf, when only the region in the vicinity of the barrier is important. However, the M3Y double-folded potential is not taking into account two major factors — the density dependence of the *NN* interaction and the Pauli principle, which are important at distances corresponding to the overlap of the nuclear volumes. This potential is characterized by a strong, unphysical attraction of a few thousands of MeV inside the nucleus. To accommodate a molecular model with the potential used in the calculations one need a repulsive core which would prevent the reabsorbtion of the lighter fragment by the heavier one. A double folding potential based on the effective Skyrme interaction is a good choice for a decaying giant molecule or dinuclear system [15,16], in view of its similarities with the interatomic potentials used in the physics of the molecule [17]. Thus the nuclear potential between two heavy ions contains an attractive part and a repulsive one. Neglecting the spin dependence, it can be written as

$$V_N(\mathbf{R}) = C_0 \left\{ \frac{F_{\text{in}} - F_{\text{ex}}}{\rho_{00}} ((\rho_1^2 * \rho_2)(\mathbf{R}) + (\rho_1^* \rho_2^2)(\mathbf{R})) \right.$$
$$\left. + F_{\text{ex}}(\rho_1^* \rho_2)(\mathbf{R}) \right\}, \quad (2)$$

where $*$ denotes the convolution of two functions $f$ and $g$, i.e., $(f*g)(\mathbf{x}) = \int f(\mathbf{x}')g(\mathbf{x}-\mathbf{x}')d\mathbf{x}'$. The constant $C_0$ and the dimensionless parameters $F_{\text{in}}, F_{\text{ex}}$ are given in Ref. [15]. To solve this integral we consider the inverse Fourier transform

$$V_N(\mathbf{R}) = \int e^{-i\mathbf{q}\cdot\mathbf{R}} \widetilde{V}_N(\mathbf{q}) d\mathbf{q}, \quad (3)$$

where the Fourier transform of the local Skyrme potential $\widetilde{V}_N(\mathbf{q})$ can be casted in the form

$$\widetilde{V}_N(\mathbf{q}) = C_0 \left\{ \frac{F_{\text{in}} - F_{\text{ex}}}{\rho_{00}} (\widetilde{\rho_1^2}(\mathbf{q})\widetilde{\rho}_2(-\mathbf{q}) + \widetilde{\rho}_1(\mathbf{q})\widetilde{\rho_2^2}(-\mathbf{q})) \right.$$
$$\left. + F_{\text{ex}}\widetilde{\rho}_1(\mathbf{q})\widetilde{\rho}_2(\mathbf{q}) \right\}. \quad (4)$$

Here $\widetilde{\rho}(\mathbf{q})$ and $\widetilde{\rho^2}(\mathbf{q})$ are Fourier transforms of the nucleon densities $\rho(\mathbf{r})$ and squared nuclear densities $\rho^2(\mathbf{r})$. Expanding the nucleon densities for axial-symmetric distributions in spherical harmonics we get

$$\rho(\mathbf{r}) = \sum_\lambda \rho_\lambda(r) Y_{\lambda 0}(\theta, 0). \quad (5)$$

Then

$$\widetilde{\rho}(\mathbf{q}) = 4\pi \sum_\lambda i^\lambda Y_{\lambda 0}(\theta_q, 0) \int_0^\infty r^2 dr \rho_\lambda(r) j_\lambda(qr), \quad (6)$$

$$\widetilde{\rho^2}(\mathbf{q}) = \sqrt{4\pi} \sum_\lambda \frac{i^\lambda}{\hat{\lambda}} Y_{\lambda 0}(\theta_q, 0)$$
$$\times \sum_{\lambda'\lambda''} \hat{\lambda}'\hat{\lambda}''(C_{0\ 0\ 0}^{\lambda\lambda'\lambda''})^2 \int_0^\infty r^2 dr \rho_{\lambda'}(r)\rho_{\lambda''}(r) j_\lambda(qr). \quad (7)$$

Like in previous papers we take the one-body densities for both daughter and cluster as two-parameter Fermi distributions in the intrinsic frame for axial symmetric nuclei

$$\rho(\mathbf{r}) = \frac{\rho_{00}}{1 + \exp([r - R(\theta)]/a)}. \quad (8)$$

Here $\rho_{00} = 0.17$ fm$^{-3}$ and the diffusivity $a$ is taken to be 0.5 fm for both fission fragments. We consider that the nuclei which compose the giant molecule are in their ground state with known quadrupole $\beta_2$ and hexadecupole deformations $\beta_4$.

The most favorable configuration which leads to decay is the one in which the fragments symmetry axes are aligned. Moreover, we constrain the fragments to rotate only around an axis perpendicularly to the axis joining their centers. Then, the only angular collective variables left are $\theta_1$ and $\theta_2$, i.e., the angles between the symmetry axes of the deformed fragments and the fission axis. This assumption is justified experimentally by the small forward anisotropy of the angular distribution of prompt $\gamma$ radiation.

Together with the above approximations we consider also that the nuclei, building-up the molecule, does not perform $\beta$ or $\gamma$ vibrations. This assumption is based on the fact that before scission the interfragment distance $R$ is rather an ellongation coordinate which describes the stretching of the whole molecule.

The classical Hamiltonian function of the giant molecule is taken in the form

$$H = \frac{\mathbf{L}_{\text{rel}}^2}{2\mu R_{12}^2} + H_{\text{coll}} + H_{\text{int}}, \quad (9)$$

where the first term represents the rotation of the whole molecule and the second,

$$H_{\text{coll}} = \frac{\mu \dot{\mathbf{R}}_{12}^2}{2} + T_{\text{rot}}(\theta_1, \theta_2) + V(\mathbf{R}_{12}, \theta_1, \theta_2), \quad (10)$$

describes the dynamics of the radial (stretching-fission) and rotational collective variables. In the last formula $T_{\text{rot}}$ is the kinetic rotational energy of the fragments. Using the multi-polar formalism presented in [18], the total (Coulomb +nuclear) heavy-ion potential (1) reads

$$V(\mathbf{R}_{12}) = \sum_{\lambda_1,\lambda_2,\lambda_3,\mu} V^{\mu-\mu 0}_{\lambda_1\lambda_2\lambda_3}(R_{12}) Y_{\lambda_1\mu}(\theta_1,0) Y_{\lambda_2-\mu}(\theta_2,0). \tag{11}$$

For small nonaxial fluctuations(bendings), the potential in the neighborhood of the scission, or ''molecular equilibrium'' point $R_{12}=R_c$, gets a simplified form, provided we keep terms up to the second power in angle:

$$V(R_c,\theta_1,\theta_2) = V(R_c,0,0) + \frac{1}{2}C_1\theta_1^2 + \frac{1}{2}C_2\theta_2^2 + C_{12}\theta_1\theta_2. \tag{12}$$

Here $C_1$ and $C_2$ are the fragment bending stiffness, and $C_{12}$ is the coupling constant. Explicitly, they are given by the relations

$$C_1 = -\frac{1}{2}\sum_{\lambda_1\lambda_2\lambda_3} \lambda_1(\lambda_1+1) V^{000}_{\lambda_1\lambda_2\lambda_3}(R_c), \tag{13}$$

$$C_2 = -\frac{1}{2}\sum_{\lambda_1\lambda_2\lambda_3} \lambda_2(\lambda_2+1) V^{000}_{\lambda_1\lambda_2\lambda_3}(R_c), \tag{14}$$

$$C_{12} = -\frac{1}{4}\sum_{\lambda_1\lambda_2\lambda_3} \{\lambda_3(\lambda_3+1) - \lambda_1(\lambda_1+1) - \lambda_2(\lambda_2+1)\} V^{000}_{\lambda_1\lambda_2\lambda_3}(R_c). \tag{15}$$

The last term in Eq. (9), $H_{int}$, is the sum of the intrinsic Hamiltonians of the two nuclei forming the giant molecule. They include independent quasiparticle excitations and residual interactions between these quasiparticles. It was separated from the rest of the Hamiltonian by means of an adiabaticity assumption. This is supported by the fact that during fission, the molecular quantum numbers are not likely to change as the intrinsic ones.

Further, in the frame of the same Born-Oppenheimer adiabatic approximation [17], the translational (stretching) mode is separated from the rotational mode by fixing first the coordinate $R_{12}=R_c$, i.e., *nailing down* the fission motion.

Thus, separating the intrinsic and relative translational motion from Eq. (10) we are left with a Hamiltonian accounting for the molecular rotational vibrations (bending) modes:

$$H_{rv}(R_c,\theta_1,\theta_2) = T_{rot}(\theta_1,\theta_2) + \frac{\mathbf{L}^2_{rel}}{2\mu R_c^2} + V(R_c,\theta_1,\theta_2). \tag{16}$$

Since the spin of the mother nucleus, $^{252}$Cf, is zero, we have a relation between the spins of the two fragments, $\mathbf{L}_1,\mathbf{L}_2$ and the relative orbital angular momentum $\mathbf{L}_{rel}$

$$\mathbf{L}_1 + \mathbf{L}_2 + \mathbf{L}_{rel} = 0. \tag{17}$$

The fragments being constrained to rotate in the same plane, in the above expression we consider only the component of the angular momentum, perpendicular to the fission axis, i.e.,

$L_k = -i\hbar\partial/\partial\theta_k (k=1,2)$. In this way the angular momentum $L$ and the deviation $\theta$ are conjugate variables.

Putting all these together, the quantized form of our molecular Hamiltonian is casted in the following form:

$$H_{rv} = -\frac{\hbar^2}{2\mathcal{B}_1}\frac{\partial^2}{\partial\theta_1^2} - \frac{\hbar^2}{2\mathcal{B}_2}\frac{\partial^2}{\partial\theta_2^2} - \frac{\hbar^2}{\mu R_c^2}\frac{\partial^2}{\partial\theta_1\partial\theta_2} + V(R_c,\theta_1,\theta_2), \tag{18}$$

where $\mathcal{B}_{1(2)}$ are related to the inertia moments of the fragments, $\mathcal{J}_{1(2)}$, by means of the formulas

$$\mathcal{B}_{1(2)} = \frac{\mathcal{J}_{1(2)}\mu R_c^2}{\mathcal{J}_{1(2)} + \mu R_c^2}. \tag{19}$$

In what follows we omit the constant term $V(R_c,0,0)$ from the molecular Hamiltonian (18).

Introducing the following notations:

$$\omega_{1(2)} = \sqrt{\frac{C_{1(2)}}{\mathcal{B}_{1(2)}}}, \quad K_q = \frac{C_{12}}{\sqrt{\mathcal{B}_1\mathcal{B}_2}}, \quad K_p = \frac{\sqrt{\mathcal{B}_1\mathcal{B}_2}}{\mu R_c^2}, \tag{20}$$

and passing to a new set of generalized coordinates $(q,p)$

$$q_i = \sqrt{\mathcal{B}_i}\theta_i, \quad p_i = \frac{L_i}{\sqrt{\mathcal{B}_i}}, \tag{21}$$

we obtain a Hamiltonian for two one-dimensional oscillators with $qq$ and $pp$ couplings

$$H_{rv} = \frac{1}{2}p_1^2 + \frac{1}{2}p_2^2 + \frac{1}{2}\omega_1^2 q_1^2 + \frac{1}{2}\omega_2^2 q_2^2 + K_q q_1 q_2 + K_p p_1 p_2. \tag{22}$$

The above Hamiltonian can be easily led to the canonical form by means of a unitary transformation:

$$\tilde{H}_{rv} = e^{-S} H_{rv} e^S, \tag{23}$$

characterized by the exponent

$$S = i(\eta_1 q_1 p_2 + \eta_2 q_2 p_1). \tag{24}$$

Requiring that

$$\eta_1 = -\frac{K_p\omega_1^2 + K_q}{K_p\omega_2^2 + K_q}\eta_2, \tag{25}$$

we obtain the diagonal form of the Hamiltonian in the new variables $(\tilde{q},\tilde{p})$

$$\tilde{H}_{rv} = \frac{1}{2\mathcal{B}_1}\tilde{p}_1^2 + \frac{1}{2\mathcal{B}_2}\tilde{p}_2^2 + \frac{1}{2}\mathcal{B}_1\Omega_1^2\tilde{q}_1^2 + \frac{1}{2}\mathcal{B}_2\Omega_2^2\tilde{q}_2^2, \tag{26}$$

where the relation between the new frequencies $\Omega_i$ and the old ones, if we choose $\omega_1 > \omega_2$, is given by

$$\Omega^2_{1(2)} = \frac{(K_q(\omega_1^2+\omega_2^2\pm\sqrt{\Delta})+2K_p\omega_1^2\omega_2^2)(K_p(\omega_1^2+\omega_2^2\pm\sqrt{\Delta})+2K_q)}{4(K_p\omega_1^2+K_q)(K_p\omega_2^2+K_q)}, \quad (27)$$

with

$$\Delta = (\omega_1^2-\omega_2^2)^2 + 4(K_p\omega_1^2+K_q)(K_p\omega_2^2+K_q). \quad (28)$$

The eigenvalues of the Hamiltonian $\tilde{H}$ are

$$E_{n_1 n_2} = \hbar\Omega_1\left(n_1+\frac{1}{2}\right) + \hbar\Omega_2\left(n_2+\frac{1}{2}\right), \quad (29)$$

whereas the eigenfunctions will be given simply by products of one-dimensional harmonic oscillators wave functions

$$\psi_{n_1 n_2}(\tilde{q}_1, \tilde{q}_2) = \psi_{n_1}(\tilde{q}_1)\psi_{n_2}(\tilde{q}_2). \quad (30)$$

### III. ANGULAR MOMENTUM IN THE COLD FISSION

#### A. Predecay angular momentum

From Eq. (30) one can easily derive the expectation values of the angular momentum operator for different molecular vibrational states. In the present approach a molecular vibrational state is defined by a couple of h.o. quantum numbers $(n_1, n_2)$. In the case when only one of the fragments is deformed and the other is spherical one could define a single vibrational quantum number $N = n_1$ or $n_2$. Since in our treatment the angular momentum operator $L = -i(\partial/\partial\theta)$ reduces to the impulse $p$, conjugated to the angle $\theta$, the computation of its matrix elements is straightforward in the Dirac-Fock representation of the harmonic oscillator. The rms of the angular momentum for each fragment in a molecular state $(n_1, n_2)$ is given by the expression

$$\sqrt{\langle L_{1,2}^2\rangle}_{n_1,n_2} = \frac{1}{\tilde{\theta}_{1,2}}\sqrt{n_{1,2}+\frac{1}{2}}, \quad (31)$$

where $\tilde{\theta}_i = 1/\sqrt{\mathcal{B}_i\Omega_i}$

The transversal angular vibrations that we treated in the preceding section can be reduced to more simple modes like bending (butterfly) or wriggling (antibutterfly). The first one corresponds to rotations in opposite directions, whereas for the second mode, both fragments rotate in the same direction. Since the binary fission mode has a pole-pole configuration, which minimize the penetration probability, the small angles $\theta_1$ and $\theta_2$ are approximately related [6]

$$\theta_1 \approx \begin{cases} \pi - \dfrac{R_2}{R_1}\theta_2: & \text{butterfly case} \\[6pt] \dfrac{R_2}{R_1}\theta_2: & \text{antibutterfly case.} \end{cases} \quad (32)$$

The angles $\theta_{1,2}$ are measured in the anticlockwise sense with respect to the fission axis. $R_1$ and $R_2$ are the fragments radii along the symmetry axes. The stiffness and mass parameters for these two modes are reading

$$C_{b(w)} = C_1 + C_2\frac{R_1^2}{R_2^2} \mp C_{12}\frac{R_1}{R_2}, \quad (33)$$

$$\frac{1}{\mathcal{B}_{b(w)}} = \frac{1}{\mathcal{B}_1} + \frac{1}{\mathcal{B}_2} \mp \frac{1}{\mu R_c^2}\frac{R_1}{R_2}. \quad (34)$$

In all the above formulas the inertia moment of each fragment $\mathcal{J}_i (i=1,2)$ was fitted to the experimental energy of the $2_1^+$ state [19].

In order to compute $\sqrt{\langle L^2\rangle}$ we need to fix $R_c$, i.e., the interfragment distance. According to Eqs. (9) and (16) the energy release in the cold fission process is the sum of zero-point energies of the ground-state configuration, i.e., the potential energy at scission $V(R_c)$, for the radial (fission) degree of freedom, $E_{n_1=0,n_2=0}$ for the transversal angular vibrations (bending) degree of freedom, and the total excitation energy $E^*$,

$$Q = V(R_c) + E_{00} + E^*. \quad (35)$$

The excitation energy comprises the contributions coming from the higher bending states $E_{\text{rv}}^*$ and the quasiparticle excitations $E_{\text{qp}}$. Owing to the fact that the radial motion is frozen at the molecular point $R_c$, the relative motion of the nascent fission fragments, i.e., the prescission kinetic energy $TKE_{\text{pre}} = 0$. This assumption is suitable for cold fission [20]. The total excitation energy will be shared between the energies stored in deformation $E_{\text{def}}$ and the prescission excitation energy $E_s^*$. Since in the spontaneous fission the dissipated energy $E_s^* \propto TKE_{\text{pre}}$, then the small amount of excitation energy present in the system will be spent for deformation. We suppose that the excitation energy is divided proportionally to the mass of each fragment, i.e., $E_i^* = [A_i/(A_1+A_2)]E^*$ and is stored only in the deformation of the fission partners. The induced deformations of each fragment can be deduced straightforwardly if one appeals to the formula for the deformation energy in the pure liquid drop model (LDM) [21] with shell corrections computed according to the Myers and Swiatecki receipt [22]

$$E_{\text{def}}(\beta) = \Delta E_{\text{surf}} + \Delta E_{\text{coul}} + \Delta E_{\text{shell}}. \quad (36)$$

In the above formula $\Delta E = E(\beta) - E(\beta_{\text{g.s.}})$. The employed version of shell corrections $\Delta E_{\text{shell}}$, is not able to describe shapes around the secondary minima of the fragments. Since in this paper we deal with very small excitation energies, bellow the neutron threshold, the aquired fragments deformations are situated around the first minima.

TABLE I. The molecular radius $R_c$, deformations of the fragments $\beta_2$, and derived rms of their angular momenta in the molecular vibrational ground state, at total excitation energies $E^*=0$, 2, and 4 MeV for the splitting $^{148}$Ba+$^{104}$Mo. In the last two columns we give the contribution to the final angular momenta due to the coulex in the case when at the starting of the trajectory calculation the fragments rotate in the same sense and are deflected at a positive angle with respect to the fission axis.

| Splitting | $E^*$ (MeV) | $R_c$ (fm) | $\beta_2(^{148}\text{Ba})$ | $\beta_2(^{104}\text{Mo})$ | $\sqrt{\langle L_1^2 \rangle}_{00}$ ($\hbar$) | $\sqrt{\langle L_2^2 \rangle}_{00}$ ($\hbar$) | $\Delta L_1$ ($\hbar$) | $\Delta L_2$ ($\hbar$) |
|---|---|---|---|---|---|---|---|---|
| $^{148}$Ba+$^{104}$Mo | 0 | 13.29 | 0.236 | 0.349 | 4.76 | 4.62 | 0.21 | 0.31 |
| $Q=211.25$ | 2 | 13.83 | 0.327 | 0.393 | 4.93 | 4.63 | 0.27 | 0.35 |
|  | 4 | 14.14 | 0.380 | 0.427 | 5.01 | 4.64 | 0.30 | 0.38 |

Since $E_{00}$ depends parametrically on $R_c$, solving the nonlinear equation (35) allows us to establish the *molecular equilibrium* point. The values of $\sqrt{\langle L^2 \rangle}_{00}$ are listed in Table I for different excitation energies.

### B. Postdecay angular momentum

After scission, the rotation of fragments is mainly influenced by the Coulomb excitation (coulex) [23]. In the case of hot fission, which does not concern the present study, the neutron emission will have a non-negligible effect on the fragments spins. In average each neutron reduces the fragment angular momenta by $0.5\hbar$.

The link between the stage when the fragments are still in the molecular configuration and the stage when the the coulex starts to act is given by the tunneling process through the barrier. In the first stage the system is found on the left side of the barrier, whereas in the second, on the right one. We suppose that between these two stages no dissipation takes place and consequently, the spins of the fragments will not be affected by the motion in the classically forbidden region. Once they arrive on the right side of the barrier the fragments will be subjected to the mutual Coulomb forces. The dynamics of the system will be given by the same Hamiltonian like in the molecular stage, see Eq. (10), with the difference that the nuclear forces are no longer contributing to the potential. We adopt a quasiclassical method to determine the asymptotic behavior of the rotational degrees of freedom [24]. First we have to solve a set of 6 first-order differential equations describing the Hamiltonain dynamics of the two fragments:

$$\dot{Q}_i = \frac{\partial H}{\partial P_i}, \quad \dot{P}_i = -\frac{\partial H}{\partial Q_i}, \quad (37)$$

where the generalized coordinates $Q_i(t)$ are given by the interfragment distance $R$, and the angular deviations of the fragments from the fission axis $\theta_{1,2}$, whereas the generalized momenta $P_i(t)$ are the translational impulse $P_R$ and the angular momenta of the fragments $L_{\theta_{1,2}}$. One needs also to establish the set of intial conditions for the above system. For $R$ we take the second turning point, i.e., the fragments relative position for which the effective decay energy $Q^*=Q-E^*$ intersects the barrier. In this case we can choose $P_R$ to be zero. For the rotational variables at $t=0$ we take the results given by the molecular model at the first turning point, i.e., $L_{\theta_{1,2}}$ are given by Eq. (31), while $\theta_{1,2}$ by

$$\sqrt{\langle \theta_{1,2}^2 \rangle} = \left(n_{1,2} + \frac{1}{2}\right) \frac{\hbar}{\sqrt{\langle L_{1,2}^2 \rangle}}. \quad (38)$$

The contribution brought by the coulex, listed on the last two columns of Table I, does not exceed 10%. This was expected, because roughly, the acquired angular momentum in coulex $\Delta L$ is $\propto \sin \theta_0$ [23], and the the rms angles given by Eq. (38), are rather small in such a way that they do not change significatively the final result. A scenario in which the coulex turns out to be important is conceivable only for large values of the initial angle, i.e., for significative nonaxial distortions at scission. Such a conjecture has been evoked recently by Rasmussen *et al.* [4], who supposed that at the cold scission configuration, the rotational population has almost entirely spin zero, while the fragments symmetry axes are in average peaked off-axis at 0.5 radians. A justification for employing such a large angle was that the neck-snapping process breaks the cylindrical shape symmetry of the fissioning system and the final neck is off axis.

Notice that the values of $\Delta L_{1,2}$ from Table I were obtained supposing that at the beginning of the accelerated motion in the Coulomb field, the fragments rotate in the same direction. If they rotate in opposite direction then for one fragment an angular momentum reduction will occur.

Note also that the asymptotic values are reached after $10^{-20}$ s of postscission motion.

As one sees from Table I, the values obtained for the spins of the $^{148}$Ba/$^{104}$Mo splitting are ranging between $4.6\hbar$ and $5\hbar$ in the neutronless fission of $^{252}$Cf. The available experimental data [10] indicates a maximum average value of the spin $\langle I_\gamma \rangle$ of $4.0\pm0.5\hbar$, for the same splitting, which obviously is in a very good agreement with our result.

### IV. CONCLUSIONS

The aim of this paper was to investigate the formation of fragments angular momenta, in the frame of a cold fission model, very similar to the deformation dependent cluster model employed earlier for the description of yields distribution in the binary cold fragmentation [14]. The scission configuration was pictured as a quasibound state of a giant molecule. In this model the angular momenta is carried by the small nonaxial vibrations of the fissioning system which arise from the higher multipole components of the interaction potential. For the cold fission we consider only the contribution of the ground state of this vibrational spectrum, the first excited state being located at $\approx 5$ MeV.

The use of a molecular model for the cold fission was also motivated by the putative existence of a giant trinuclear molecular state in the $^{10}$Be-accompanied ternary fragmentation of $^{252}$Cf [25,26].

In the case of pure cold fission ($E^*=0$ MeV) the fragments deformations are taken to be those corresponding to the first minima in the deformation energy landscape. When the excitation energy increases, we recalculated the deformations by employing the LDM with a phenomenological receipt for the shell corrections. When moving to even higher excitation energies, beyond the cold fission limit, a case which does not concern the present paper, a more reliable way to include the shell effects would be to appeal to a powerful method, like the Hartree-Fock with pairing correlations included.

We should also mention that in the present paper, like in the rest of the literature, the problem of angular momentum variation when crossing the classically forbidden region was not considered. The occurrence of friction in this region could be a source of angular momentum damping [27]. In the case of cold fission one may hope that the magnitude of such forces will not affect the final result in a decisive way.

Very recently [28], it was advocated that the fragments spins at scission could be accounted based on their intrinsic nuclear states and the uncertainty relation. Although this approach has the merit of taking into account for the first time the microscopic structure of fragments in evaluating the fragments spins, it suffers of the insufficiency of including the interaction between the fragments only by means of a resulting strong space polarization along the fission axis. As we showed throughout this paper the mutual nuclear and Coulomb torques of the fission partners is the decisive factor in the formation of angular momentum in the cold fragmentation process. In the approach presented in this paper, the population of molecular states is a consequence of small non-axial fluctuations around the axial equilibrium position, whereas in the postsnapping model of Ref. [28], the angular momentum comes from the orientation fluctuations of a strongly polarized system of two rotators.

## ACKNOWLEDGMENTS

One of the authors (Ş.M.) would like to express his gratitude to Professor I.N. Mikhailov, Professor P. Quentin, Professor F. Gönnenwein, Dr. A. Nasirov, Dr. M. Mutterer, and Dr. F. Carstoiu for illuminating discussions during the completion of this work.